\documentclass[prd,nofootinbib,preprint,superscriptaddress]{revtex4}

\usepackage{amsmath}
\usepackage[dvips]{graphicx}

\newcommand{\be}{\begin{equation}}
\newcommand{\ee}{\end{equation}}
\newcommand{\bea}{\begin{eqnarray}}
\newcommand{\eea}{\end{eqnarray}}

\newcommand{\capdef}{}
\newcommand{\mycaption}[2][\capdef]{\renewcommand{\capdef}{#2}%
       \caption[#1]{{\footnotesize #2}}}
\makeatletter
\renewcommand{\fnum@table}{\textbf{\tablename~\thetable}}
\renewcommand{\fnum@figure}{\textbf{\figurename~\thefigure}}
\makeatother

\newcommand{\Dmq}{\Delta m^2}

\newcommand{\bet}{\ensuremath{|\beta_{\mu e}|}}
\newcommand{\alp}{\ensuremath{|\alpha_{\mu e}|}}
\newcommand{\betl}{\ensuremath{\beta_{\mu e}^{\rm LK}}}
\newcommand{\alpl}{\ensuremath{\alpha_{\mu e}^{\rm LK}}}

\begin{document}


\vspace*{10mm}

\title{MiniBooNE and LSND data: non-standard neutrino
interactions in a (3+1) scheme versus (3+2) oscillations}

\author{Evgeny Akhmedov} 
\email{akhmedov@mpi-hd.mpg.de}
\affiliation{Max-Planck-Institut f\"ur Kernphysik, PO Box 103980, 
69029 Heidelberg, Germany}
\affiliation{National Research Center ``Kurchatov Institute''\\
123182 Moscow, Russia \vspace*{1.2cm} }

\author{Thomas Schwetz}
\email{schwetz@mpi-hd.mpg.de}
\affiliation{Max-Planck-Institut f\"ur Kernphysik, PO Box 103980, 
69029 Heidelberg, Germany}

\vspace*{0.5cm}

\begin{abstract}
\vspace*{.5cm} The recently observed event excess in MiniBooNE anti-neutrino
data is in agreement with the LSND evidence for electron anti-neutrino
appearance.  We propose an explanation of these data in terms of a (3+1)
scheme with a sterile neutrino including non-standard neutrino interactions
(NSI) at neutrino production and detection. The interference between
oscillations and NSI provides a source for CP violation which we use to
reconcile different results from neutrino and anti-neutrino data.  Our best
fit results imply NSI at the level of a few percent relative to the standard
weak interaction, in agreement with current bounds. We compare the quality
of the NSI fit to the one obtained within the (3+1) and (3+2) pure
oscillation frameworks. We also briefly comment on using NSI (in an
effective two-flavour framework) to address a possible difference in
neutrino and anti-neutrino results from the MINOS experiment.
\end{abstract}

\preprint{EURONU-WP6-10-24}

\maketitle


\section{Introduction}

Recently the MiniBooNE collaboration announced updated results of their
search for $\bar\nu_\mu\to\bar\nu_e$
transitions~\cite{AguilarArevalo:2010wv}. In the full energy range from
200~MeV to 3~GeV they find an excess of $43.2\pm22.5$ events over expected
background (the error includes statistical and systematical uncertainties).
In the oscillation-sensitive region of 475~MeV to 1250~MeV the
background-only-hypothesis has a probability of only
0.5\%~\cite{AguilarArevalo:2010wv}. This result is consistent with the 
evidence for $\bar\nu_\mu \to \bar\nu_e$ transitions reported by 
LSND~\cite{Aguilar:2001ty}. Any explanation of these hints for $\bar\nu_\mu
\to \bar\nu_e$ transitions at the scale of $E/L \sim 1$~eV$^2$ has to
satisfy strong constraints from various experiments. First, no evidence for
transitions has been found in MiniBooNE neutrino data above
475~MeV~\cite{AguilarArevalo:2007it}. This suggests that CP (or even CPT)
violation has to be invoked to reconcile neutrino and anti-neutrino data.
Second, severe constraints exist for $\bar\nu_e$ \cite{Declais:1994su,
Apollonio:2002gd} and $\nu_\mu, \bar\nu_\mu$ \cite{Dydak:1983zq,
AguilarArevalo:2009yj, Ashie:2005ik, Bilenky:1999ny} disappearance at this
scale, which have to be respected by any explanation of the
$\bar\nu_\mu\to\bar\nu_e$ excesses. 

The standard approach to the LSND problem is to introduce one or more
sterile neutrinos at the eV scale~\cite{Peltoniemi:1993ec,
Peltoniemi:1992ss, Caldwell:1993kn}. Adding one sterile neutrino one obtains
the so-called (3+1) mass scheme. In this framework there is no CP violation
at short baselines, and disappearance experiments strongly disfavour an 
explanation of the appearance signals, see for example~\cite{Strumia:2002fw,
Maltoni:2002xd}. If two neutrino mass states at the eV scale are present 
\cite{Peres:2000ic, Sorel:2003hf} ((3+2) scheme), the possibility of CP
violation opens up~\cite{Karagiorgi:2006jf}, which allows to reconcile LSND
and MiniBooNE neutrino data~\cite{Maltoni:2007zf}. However, constraints from
disappearance data still impose a challenge to the fit, and the overall
improvement with respect to the (3+1) case is not 
significant~\cite{Maltoni:2007zf,Karagiorgi:2009nb}. In the following we 
will update the (3+2) results of~\cite{Maltoni:2007zf} with respect to the
latest data from MiniBooNE anti-neutrinos.

Apart from sterile neutrino oscillations, various more exotic explanations
of the LSND signal have been proposed, among them, neutrino
decay~\cite{Ma:1999im, PalomaresRuiz:2005vf}, CPT
violation~\cite{Murayama:2000hm, Barenboim:2002ah, GonzalezGarcia:2003jq,
Barger:2003xm, Strumia:2002fw}, violation of Lorentz
symmetry~\cite{Kostelecky:2004hg, deGouvea:2006qd, Katori:2006mz}, quantum
decoherence~\cite{Barenboim:2004wu, Farzan:2008zv}, mass-varying
neutrinos~\cite{Kaplan:2004dq, Barger:2005mh}, shortcuts of sterile
neutrinos in extra dimensions~\cite{Pas:2005rb} or sterile neutrino
oscillations with a non-standard energy dependence~\cite{Schwetz:2007cd}. 

In this work we attempt an explanation of the global data by departing from a
(3+1) neutrino scheme through the inclusion of non-standard
interactions (NSI) of neutrinos on top of the Standard Model weak
interactions.  Such new interactions may be induced by generic new physics
beyond the Standard Model and can be neutral current (NC) like (involving
only neutrinos in the lepton sector) or charged current (CC) like (involving 
a charged lepton and a neutrino). Model-independent bounds on such new 
interactions are at the level of few~$\times 10^{-2}$ compared to the 
standard 4-Fermi interaction strength set by $G_F$, see~\cite{Biggio:2009nt}
and references therein. An observation of NSI at that level would be a 
remarkable sign of new physics. The realization of NSI in terms of effective 
operators has been discussed recently in \cite{Antusch:2008tz,Gavela:2008ra}.

Since the experiments considered here typically have rather short baselines
(below 1~km), matter effects are very small and NSI affecting the
propagation of neutrinos through matter will have a negligible impact.
Therefore, we focus on CC-like NSI in the production and detection
processes, see for example~\cite{Grossman:1995wx, GonzalezGarcia:2001mp,
Huber:2002bi, Meloni:2009ia}. As pointed out in \cite{GonzalezGarcia:2001mp},
such NSI provide a new source of CP violation. Here we use this effect to
reconcile neutrino and anti-neutrino data by exploring the interference of
the NSI with sterile neutrino oscillations, with $\Delta m^2$ of order
1~eV$^2$. This allows us to keep NSI at the percent level. Previous attempts
to explain the LSND excess by NSI without sterile neutrinos can be found
in~\cite{Bergmann:1998ft, Bergmann:2000gn, Babu:2002ica}, while sterile
neutrinos with an exotic matter effect have been considered
in~\cite{Zurek:2004vd, Nelson:2007yq}.

In the following section~\ref{sec:II} we derive an effective
parameterisation of the transition and survival probabilities for
short-baseline experiments, identifying the particular combinations of
mixing and NSI parameters relevant for the experiments. We show that the
probabilities for the global short-baseline data in a general (3+1) NSI
model depend only on 8 independent parameters. In this case we can make use
of the fact that in LSND and KARMEN the neutrino production mechanism is
different from all other experiments. Therefore these experiments decouple
to some extent from the rest of the global data. 
On the other hand, neglecting NSI involving the charged muon, only 5
independent parameters remain, which are only two independent parameters
more than in the case of standard (3+1) oscillations, one modulus and one
phase. In this case the fit is more constrained, while still providing
significant improvement with respect to the pure oscillation case. As we
will see, this constrained version requires only one single non-zero NSI
parameter. In section~\ref{sec:NSI} we present the results of global fits
within these two versions of the (3+1) NSI model, the constrained 5
parameter version in section~\ref{sec:NSIc} and the general one in
\ref{sec:NSIg}. Furthermore, in section~\ref{sec:3+2} we present an updated
fit in the (3+2) oscillation scheme and we compare the quality of the (3+1)
NSI fits to the one of (3+2) oscillations. A summary and discussion follow
in section~\ref{sec:summary}. In appendix~\ref{sec:vector} we relax some of
the assumptions made on NSI parameters (showing that they do not change our
results significantly), and in appendix~\ref{sec:minos} we comment briefly
on using NSI to address a possible difference in neutrino and anti-neutrino
results from the MINOS experiment~\cite{minos:an}.

\section{NSI in the (3+1) mass scheme}
\label{sec:II}

\subsection{The formalism} 
\label{sec:formalism}

We assume that, in addition to the standard charged current (CC) weak 
interactions, there exist non-standard CC-like interactions, whose 
Lagrangian can be parameterised at low energies as  
\be
{\cal L}_{\rm NSI}=-2\sqrt{2}G_F\, \sum_{\alpha,\beta}
\varepsilon^{ff'}_{\alpha\beta}(\bar{f} P_{L,R} \gamma^\mu f')
(\bar{l}_{\alpha} P_L \gamma_\mu \nu_{\beta})+h.c.\,.
\label{eq:eps}
\ee
Here $G_F$ is the Fermi constant, $f$ and $f'$ correspond to either quarks
or leptons differing by one unit of electric charge and $l_{\alpha}$
corresponds to a charged lepton ($l_{\alpha}=e, \mu, \tau$). $P_{L(R)}$
denotes the projection operator on left-handed (right-handed) fields. The
particular chirality structure assumed in eq.~(\ref{eq:eps}) allows for
interference of standard and non-standard processes. To simplify the
notation we leave the dependence of $\varepsilon$ on $L/R$ implicit. 
  
The interactions in eq.~(\ref{eq:eps}) contribute to CC processes of 
neutrino emission and absorption. In what follows, we will replace the 
superscript $ff'$ by $X$ ($X=S,D$), where $X=S$ stands for a concrete neutrino 
production process in the source and $X=D$ stands for the neutrino detection 
process. 
In the presence of NSI the neutrino state $|\nu_\alpha^X\rangle$ produced or 
detected in a CC process along with a charged lepton $l_\alpha$ is no longer 
the flavour eigenstate $\nu_\alpha$, but a linear combination of flavour 
eigenstates:
\be
|\nu_\alpha^X\rangle=
C_\alpha^X\Big(|\nu_\alpha\rangle+\sum_\beta
\varepsilon_{\alpha\beta}^X|\nu_\beta\rangle\Big). 
\label{eq:nu1}
\ee
A similar relation holds for anti-neutrinos, with
$\varepsilon_{\alpha\beta}^X$ replaced by $\varepsilon_{\alpha\beta}^{X*}$.
Note that in our notation the first subscript at the parameter
$\varepsilon^X_{\alpha\beta}$ always refers to the charged lepton and the
second to the neutrino.
$C_\alpha^X$ is a normalisation constant, which satisfies 
\be
|C_\alpha^X|^2\Big(1+2{\rm Re}\,\varepsilon_{\alpha\alpha}^X+\sum_\rho
|\varepsilon_{\alpha\rho}^X|^2 \Big)=1\,. \label{eq:norm} 
\ee 
We also assume that, in addition to the standard three neutrino flavour
states $\nu_e$, $\nu_\mu$ and $\nu_\tau$, there exists a fourth light
neutrino, which necessarily has to be a sterile neutrino $\nu_s$. As usual,
the flavour states $\nu_e$, $\nu_\mu$, $\nu_\tau$ and $\nu_s$ are linear
combinations of four mass eigenstates:\footnote{In the presence of NSI
given in eq.~(\ref{eq:eps}), the usual flavour states can only be defined at
the level of the Lagrangian, as the states coupled to the Standard Model
$W$ boson. Therefore, the mixing matrix $U$ is the unitary matrix which
diagonalises the neutrino mass matrix in the basis where the Standard Model
CC interaction (without new physics contributions) is diagonal.}
\be 
|\nu_\alpha\rangle = \sum_i U_{\alpha i}^* |\nu_i\rangle\,.
\label{eq:mix} 
\ee 
We assume that the 4th mass eigenstate is separated from the other three by
a mass gap of order $\Delta m_{41}^2\sim 1$ eV$^2$ ((3+1) scheme).
Alternatively, we will also consider a variant with two sterile neutrinos
but no NSI ((3+2) scheme). 

The state $|\nu_\alpha^X\rangle$ can also be expressed in terms of the 
mass eigenstates:
\be
|\nu_\alpha^X\rangle\,=\,C_\alpha^X\sum_i\Big(U_{\alpha i}^*+
\sum_\beta \varepsilon_{\alpha\beta}^X U_{\beta i}^*\Big)|\nu_i\rangle 
\,=\,
C_\alpha^X 
\sum_{\beta,i}\left(\delta_{\alpha\beta}+\varepsilon_{\alpha\beta}^X\right)
U_{\beta i}^*|\nu_i\rangle\,.
\label{eq:nu2}
\ee
The probability amplitude for a neutrino $\nu_\alpha^S$ born in a 
production process in the neutrino source to be detected as $\nu_\beta^D$ 
at a distance $L$ from the source in a detection process is 
\be
{\cal A}_{\alpha\beta}(L)=\langle \nu_\beta^D|\nu_\alpha^S(L)\rangle=
\sum_i F_{\alpha i}^{S} F_{\beta i}^{D*} e^{-i E_i L}\,,
\label{eq:amp1}
\ee 
where $E_i$ is the energy of the $i$th mass eigenstate neutrino and 
\be
F_{\alpha i}^X \equiv C_\alpha^X \sum_\rho \left(\delta_{\alpha\rho}+
\varepsilon_{\alpha\rho}^{X}\right)\!U_{\rho i}^*\,.
\label{eq:F1}
\ee 

Next, we note that for baselines and neutrino energies of interest, one 
can neglect all mass squared differences except those involving the fourth 
mass eigenstate. Denoting 
\be
\Delta \equiv \frac{\Delta m_{41}^2}{2E} L\,,
\label{eq:Delta}
\ee 
we can rewrite eq.~(\ref{eq:amp1}) in the generic form
\be
{\cal A}_{\alpha\beta}(L)=\alpha_{\alpha\beta}(e^{-i \Delta}-1)+
\beta_{\alpha\beta}\,,
\label{eq:amp2}
\ee
where 
\be
\alpha_{\alpha\beta}=
F_{\alpha 4}^{S} F_{\beta 4}^{D*}\,,\qquad\quad
\beta_{\alpha\beta} = \sum_i F_{\alpha i}^{S} F_{\beta i}^{D*}\,.
\label{eq:F2}
\ee
The transition probability is then the squared modulus of the amplitude 
(\ref{eq:amp2}): 
\be
P_{\alpha\beta}(L)=
2\left[|\alpha_{\alpha\beta}|^2
-{\rm Re}(\beta_{\alpha\beta}^*\alpha_{\alpha\beta})\right](1-\cos \Delta)
+|\beta_{\alpha\beta}|^2+
2 {\rm Im}(\beta_{\alpha\beta}^*\alpha_{\alpha\beta}) \sin \Delta\,.
\label{eq:P1}
\ee
The probabilities for anti-neutrinos can be obtained from this formula by
complex-conjugating the parameters $\alpha_{\alpha\beta}$ and 
$\beta_{\alpha\beta}$, i.e.\ by flipping the sign of the last term in
eq.~(\ref{eq:P1}). Thus, in the presence of NSI the transition probabilities
exhibit CP violation even in the one mass scale dominance limit, unlike in
the case of pure oscillations. We will exploit this property of the
considered model in order to reconcile the neutrino and anti-neutrino data
in short-baseline appearance experiments. Since CP violation comes through
the interference of $\alpha_{\alpha\beta}$ and $\beta_{\alpha\beta}$ terms
it is important that both terms be present. It follows from
eq.~(\ref{eq:P1}) that in order to suppress (enhance) the transition
probability for neutrinos (anti-neutrinos) we need the phase of
$\beta_{\alpha\beta}^*\alpha_{\alpha\beta}$ to be close to $3\pi/2$. Indeed,
in our fits to be discussed in the following we will obtain numbers close to
this value.
Note that in general $P_{\alpha\beta}$ summed over either of the two indices
will not give unity. This is because NSI break the unitarity of the neutrino
evolution~\cite{FernandezMartinez:2007ms}.

\subsection{Application to short-baseline data} 
\label{sec:params}

\begin{table}[t] \centering
  \begin{tabular}{l@{\quad}c@{\quad}c@{\quad}r@{\quad}
                   @{\quad}l@{\quad}c@{\quad}c@{\quad}r}
	\hline\hline
	\multicolumn{4}{c}{Disappearance} & \multicolumn{4}{c}{Appearance} \\
	\hline
        Experiment & Ref. & Channel & Data & Experiment & Ref. & Channel & Data \\
	\hline
	Bugey     & \cite{Declais:1994su}  &$\bar\nu_e\to\bar\nu_e$ & 60 & 
	LSND      & \cite{Aguilar:2001ty}  &$\bar\nu_\mu\to\bar\nu_e$& 11 \\
	Chooz     & \cite{Apollonio:2002gd}&$\bar\nu_e\to\bar\nu_e$ &  1 &
        KARMEN    & \cite{Armbruster:2002mp}&$\bar\nu_\mu\to\bar\nu_e$&  9 \\
	Palo Verde& \cite{Boehm:2001ik}    &$\bar\nu_e\to\bar\nu_e$ &  1 &
        NOMAD     & \cite{Astier:2003gs}   &$\nu_\mu\to\nu_e$       &  1 \\
        CDHS      & \cite{Dydak:1983zq}    &$\nu_\mu\to\nu_\mu$     & 15 &
        MiniB ($\nu$) 
	          & \cite{AguilarArevalo:2007it} &$\nu_\mu\to\nu_e$       &  8 \\
	atmospheric& \cite{Maltoni:2007zf}  &$\nu_\mu\to\nu_\mu$     &  1 &
        MiniB ($\bar\nu$) & \cite{AguilarArevalo:2010wv} &$\bar\nu_\mu\to\bar\nu_e$ &  8 \\
	\hline\hline
    \end{tabular}
    \mycaption{\label{tab:experiments} Experiments used in the numerical
     analysis. The oscillation channel and the number of data points
     used in the fit (``Data'') are given. The total number of data points
     is 115.}
\end{table}

Let us now identify the effective parameters which are relevant for describing
the short-baseline data used in our analysis, as summarised in
tab.~\ref{tab:experiments}, where we divide the experiments into
appearance and disappearance searches.

\bigskip

1. Appearance experiments ($\mu\to e$). In this case we have
\be
\alpha_{\mu e}=F_{\mu 4}^S F_{e 4}^{D*}\,,
\label{eq:mue1}
\ee
where
\bea
F_{\mu 4}^S &=&
C_\mu^S \left(U_{\mu 4}^*+\varepsilon_{\mu e}^S U_{e4}^*+\varepsilon_{\mu\mu}^S
U_{\mu4}^*+\varepsilon_{\mu\tau}^S U_{\tau4}^*+\varepsilon_{\mu s}^S 
U_{s4}^*\right),
\nonumber \\
F_{e 4}^D &=& 
C_e^D\left(U_{e4}^*+\varepsilon_{ee}^D U_{e4}^*+\varepsilon_{e\mu}^D U_{\mu4}^*
+\varepsilon_{e\tau}^D U_{\tau4}^*+\varepsilon_{es}^D U_{s4}^*\right).
\label{eq:FeFmu}
\eea
For the parameter $\beta_{\mu e}$ we obtain 
\be \beta_{\mu e}=\sum_i F_{\mu i}^S F_{ei}^{D*}= C_\mu^S
C_e^{D*}\Big(\varepsilon_{\mu e}^S+\varepsilon_{e\mu}^{D*}+ \sum_\rho
\varepsilon_{\mu\rho}^S\varepsilon_{e\rho}^{D*}\Big). \label{eq:betamue} 
\ee
In deriving this expression from eqs.~(\ref{eq:F2}) and (\ref{eq:F1}) we
have used unitarity of the leptonic mixing matrix. Notice that the 
transition probability $P_{\mu e}(L)$ does not vanish in the limit $L=0$: in
that case $P_{\mu e}=|\beta_{\mu e}|^2$. This is related to the fact that in
the presence of NSI the states $|\nu_\mu^S\rangle$ and $|\nu_e^D\rangle$ are
not orthogonal~\cite{Langacker:1988up}, see eq.~(\ref{eq:nu1}). This should
lead to a nontrivial signal in near detectors in appearance experiments.  

\bigskip

2. Disappearance experiments.  We will need the survival probabilities
$P_{ee}$ and $P_{\mu\mu}$. For the processes of interest (pion decay,
nuclear beta decay and inverse beta processes), both neutrino production and
detection involve transitions between $u$ and $d$ quarks. For the reactor
experiments relevant for $P_{ee}$, production and detection are just inverse
processes to each other. Therefore the relevant NSI are identical. For
$P_{\mu\mu}$ the production process is pion decay, whereas detection
proceeds via neutrino--nucleon capture. Since pions couple only to the
axial current, vector-like NSI would not contribute to neutrino production
but still show up in detection.\footnote{We would like to thank Joachim Kopp
for drawing our attention to this point.} Hence, in this case it is possible
that NSI at production and detection are different. We will discuss this
possibility in appendix~\ref{sec:vector}. For the moment we simplify the
discussion and assume that also for $\nu_\mu$ disappearance NSI are
identical in source and detection:
\be\label{eq:source-det}
\varepsilon_{\mu\alpha}^S =\varepsilon_{\mu\alpha}^D \qquad
\text{($\nu_\mu$ disappearance)}\,.
\ee
This corresponds to axial-vector like NSI, $\varepsilon^{(A)} =
\varepsilon^{(R)} - \varepsilon^{(L)}$. In this case we find from
eqs.~(\ref{eq:F2}) and (\ref{eq:norm}) 
\be 
\alpha_{ee}\equiv \alpha_e=|F_{e4}^{ud}|^2\,, \qquad 
\alpha_{\mu\mu}\equiv \alpha_\mu=|F_{\mu 4}^{ud}|^2\,, \qquad 
\beta_{ee}=\beta_{\mu\mu}=1. \label{eq:eemumu} 
\ee 
The equality of $\beta_{ee}$ and $\beta_{\mu\mu}$ to unity is a reflection
of the fact that at $L=0$ the survival probabilities are equal to one. It
should be stressed that this is only correct in the case when the neutrino
production and detection processes are of the same type, i.e., 
$\varepsilon_{\alpha\beta}^S =\varepsilon_{\alpha\beta}^D$, as in the case
we consider here. Otherwise the survival probabilities at $L=0$ would be smaller
than one.  
%
%
From eqs.~(\ref{eq:P1}) and (\ref{eq:eemumu}) we find for the survival 
probabilities
\be
P_{\beta\beta} = 1 - 2 \alpha_\beta (1 - \alpha_\beta) (1 - \cos\Delta) 
\,,\qquad \beta = e,\mu\,.
\label{eq:Pbetabeta}
\ee
There are no CP violation effects for the survival probabilities, again
because we take $\varepsilon_{\alpha\beta}^S =\varepsilon_{\alpha\beta}^D$.
Note that eq.~(\ref{eq:Pbetabeta}) has the same structure as the
disappearance probability for (3+1) oscillations with the identification
$\alpha_\beta \to |U_{\beta 4}|$, see e.g.~\cite{Bilenky:1996rw}.

In appendix~\ref{sec:vector} we will relax the assumption
(\ref{eq:source-det}) and allow $\varepsilon_{\alpha\beta}^S \neq
\varepsilon_{\alpha\beta}^D$ for $\nu_\mu$ disappearance experiments. As we
will discuss there, assuming $\varepsilon_{\alpha\beta}^S
=\varepsilon_{\alpha\beta}^D$ has a small impact on the global fit and
therefore we will adopt this simplification for our discussion apart from
appendix~\ref{sec:vector} and where stated explicitly.

\bigskip

As mentioned above, for the disappearance experiments neutrino
production and detection processes involve transitions between $u$ and
$d$ quarks. This holds also for the MiniBooNE and NOMAD appearance
experiments. The only exceptions are LSND and KARMEN. While the
detection process still involves $u$ and $d$ quarks (inverse beta
decay), the production process is purely leptonic (muon
decay)\footnote{We not not consider here the decay-in-flight data from
LSND, where neutrinos are produced in pion decay. Muon decay
contributes also to neutrino production in atmospheric neutrinos. We
do not expect that this will have an important impact on our analysis
and neglect this effect in the following, treating all NSI effects in
atmospheric neutrinos as semi-leptonic.}, see also eqs.~(\ref{eq:a1})
and (\ref{eq:a2}) in the appendix.
Therefore, all experiments except LSND and KARMEN only depend on
semi-leptonic NSI, and (under the assumption eq.~(\ref{eq:source-det})) we
can drop the distinction between source and detector:
\be\label{eq:F_SD}
F_{\alpha i}^S = F_{\alpha i}^D \equiv F_{\alpha i}^{ud} \qquad 
\text{(semi-leptonic NSI only}) \,.
\ee
This means that the same NSI parameters appear for processes involving a given
charged lepton $\alpha$, irrespective of whether the process occurs at
neutrino production or detection. If this relation holds, we obtain from
eqs.~(\ref{eq:mue1}) and (\ref{eq:eemumu}) a factorisation property of the
parameters $\alpha$:
\be
|\alpha_{\mu e}|^2=\alpha_{e}\alpha_{\mu}\,.
\label{eq:factor}
\ee
Therefore, all experiments except LSND and KARMEN depend on the following
set of parameters:
\be\label{eq:params_q}
\alpha_e, \, \alpha_\mu, \, \bet,\,
\delta \equiv {\rm Arg}\left(\alpha_{\mu e}\beta_{\mu e}^* \right) 
, \, \Dmq_{41}\,.
\ee

For LSND and KARMEN purely leptonic NSI will also contribute, while
semi-leptonic NSI involving the charged muon, $\varepsilon_{\mu \alpha}^{ud}$,
will contribute to the other experiments but not to LSND and KARMEN. This
leads effectively to a decoupling of these experiments from the others, and
we can describe LSND and KARMEN by the three independent parameters 
\be\label{eq:params_LK}
|\alpl|, \,|\betl|, \, \delta^{LK} \equiv {\rm Arg}(\alpl {\betl}^*) 
\ee
in addition to the common $\Dmq_{41}$. Note that in this framework MiniBooNE
cannot be considered as a direct test of LSND, since due to the different
production mechanisms different parameters enter the transition
probabilities for the two experiments.

Therefore, the global data in the general (3+1) NSI scenario (under the
assumption eq.~(\ref{eq:source-det})) depend on the 8 independent parameters
shown in eqs.~(\ref{eq:params_q}) and (\ref{eq:params_LK}). We shall denote
this case by NSI$^g$. In appendix~\ref{sec:vector} we relax the
assumption eq.~(\ref{eq:source-det}) and argue that the effect on the global
fit is small. In addition to the general NSI$^g$ case we will discuss in the
following also a constrained scenario, denoted by NSI$^c$, which can be
obtained by assuming that the relation (\ref{eq:F_SD}) holds for the global
data, also including LSND and KARMEN. This can be achieved by assuming that
all NSI parameters involving the charged muon (leptonic as well as
semi-leptonic) vanish:
\be\label{eq:approx_mu}
\varepsilon_{\mu \beta}^X=0 \,.
\ee
In this case, we find from eq.~(\ref{eq:F1}) that $F_{\mu i}^S=F_{\mu
i}^D=U_{\mu i}^*$ and eq.~(\ref{eq:F_SD}) holds for all experiments. Note
that for the processes relevant to the experiments we consider we always
have $F_{ei}^S = F_{ei}^D$, and if eq.~(\ref{eq:approx_mu}) is fulfilled,
also eq.~(\ref{eq:source-det}) holds trivially. Then the global data depends
only on the 5 parameters from eq.~(\ref{eq:params_q}) and the factorisation
(\ref{eq:factor}) applies in general. 

The factorisation (\ref{eq:factor}) is analogous to the factorisation
property that holds in the pure (3+1) oscillations
case~\cite{Bilenky:1996rw}, and $4 \alp^2$ can be identified with
$\sin^22\theta_{\rm SBL}$ in the limit $\beta_{\mu e} \to 0$, $\theta_{\rm
SBL}$ being an effective two-flavour mixing angle.  Hence, for $\beta_{\mu
e} = 0$ we recover the (3+1) oscillation case.
The fact that the parameter $\alpha_{\mu e}$ of the appearance amplitude
given in eq.~(\ref{eq:factor}) is a product of two quantities which are
constrained to be small from disappearance data leads to the well-known
tension between the LSND signal and disappearance data in (3+1) schemes.
Including now NSI (still subject to the approximation (\ref{eq:approx_mu}))
we obtain two more independent parameters, namely $|\beta_{\mu e}|$ and the
phase $\delta$. Since $|\beta_{\mu e}|$ enters only in the appearance
probability and is unconstrained by disappearance data, one might expect
some improvement of the tension between these two data sets. We expect
significant improvement of the fit in the NSI$^g$ case, due to the
decoupling of the LSND evidence from the rest of the global data.

In the following section~\ref{sec:NSI} we will present the results of a fit
to data in terms of the effective parameters given in
eqs.~(\ref{eq:params_q}) and (\ref{eq:params_LK}). However in both cases,
NSI$^c$ and NSI$^g$, we will present also a specific realization in terms of
the fundamental parameters $U_{\alpha 4}$ and $\varepsilon_{\alpha\beta}$. 

\section{Numerical results}
\label{sec:NSI}

In this section we present the results of fits to short-baseline data in
the (3+1) NSI framework. The data used in the numerical analysis are
summarised in tab.~\ref{tab:experiments}. In total 115 data points are used.
Technical details on our re-analysis of the experiments can be found
in~\cite{Maltoni:2007zf} and references therein. 
The constraint from atmospheric neutrino data is implemented in the
following way. As discussed in detail in~\cite{Maltoni:2007zf}, atmospheric
neutrinos provide a constraint on the parameter $d_\mu$ defined in that
paper. $d_\mu$ corresponds to $|U_{\mu 4}|^2$ in the (3+1) oscillation case.
Using the equivalence of the expressions for the survival probability in the
(3+1) oscillation and (3+1) NSI schemes in eq.~(\ref{eq:Pbetabeta}), we
identify the parameters $\alpha_\mu$ with $d_\mu$ from~\cite{Maltoni:2007zf}
and generalise the bound from atmospheric neutrinos now to hold for the
parameter $\alpha_\mu$ in the presence of NSI. A more accurate analysis of
atmospheric neutrino data with NSI effects in source and detector is
beyond the scope of this work.
For the MiniBooNE anti-neutrino sample we fit the data shown in fig.~1 of
\cite{AguilarArevalo:2010wv}, calibrating our prediction to the histogram
shown in the lower part of that figure. As we will see later, in our model
it is not possible to explain the low energy excess seen in MiniBooNE
neutrino data (and to a lesser extent also in anti-neutrino data) below
475~MeV. Therefore, we follow the MiniBooNE collaboration and restrict the
analysis to the oscillation-sensitive region above 475~MeV.

\begin{figure}
  \includegraphics[width=0.48\textwidth]{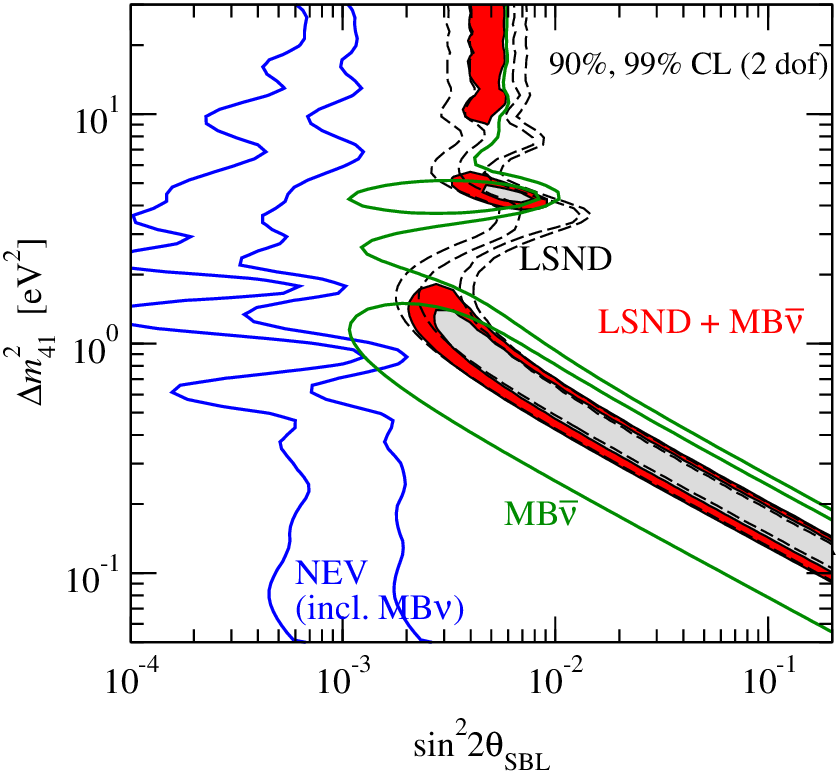}
  \mycaption{\label{fig:ne-vs-ev-osc} Constraint from no-evidence data (NEV)
  compared to the combined allowed regions from LSND and MiniBooNE $\bar\nu$
  data (shaded) at 90\% and 99\% CL for (3+1) oscillations. We show also the
  individual regions from LSND and MiniBooNE $\bar\nu$ data.}
\end{figure}

In fig.~\ref{fig:ne-vs-ev-osc} we show our fit to MiniBooNE anti-neutrino
data compared to the LSND region in the (3+1) oscillation framework without
NSI. Our result is in good agreement with the region obtained in
\cite{AguilarArevalo:2010wv}.\footnote{Differences can be attributed to the
more sophisticated statistical analysis applied in
\cite{AguilarArevalo:2010wv}, in contrast to the simple $\Delta\chi^2$
method based on the Gaussian approximation for 2~dof used here. In
particular, Ref.~\cite{AguilarArevalo:2010wv} obtains a closed contour at
99\%~CL, while we find only a limit at 99\%~CL.} We observe the overlap with
the region indicated by LSND, which motivates the combined analysis of the
two experiments shown as shaded regions. We also show the constraint coming
from all the other experiments (``no-evidence'' data, NEV), which excludes
the region to the right of the blue curves. The figure shows the well known
tension in the (3+1) oscillation fit: the regions touch each other at
$\Delta\chi^2 = 12.7$, which corresponds to 99.8\%~CL for 2~dof.

\subsection{The constrained (3+1) NSI model}
\label{sec:NSIc}

Now we proceed to the (3+1) NSI model, starting with the constrained version
NSI$^c$. As discussed in section~\ref{sec:params}, in this case the global
data depends on the 5 parameters from eq.~(\ref{eq:params_q}): $\alpha_e$,
$\alpha_\mu$, $|\beta_{\mu e}|$, $\delta$, $\Delta m^2_{41}$.  For
appearance data alone the parameters $\alpha_e$ and $\alpha_\mu$ enter only
in the combination $\alp \equiv \sqrt{\alpha_e\alpha_\mu}$, while
disappearance data do not depend on \bet\ and $\delta$. The limit of (3+1)
oscillations is obtained for $\bet = 0$. 

\begin{figure}
  \includegraphics[width=\textwidth]{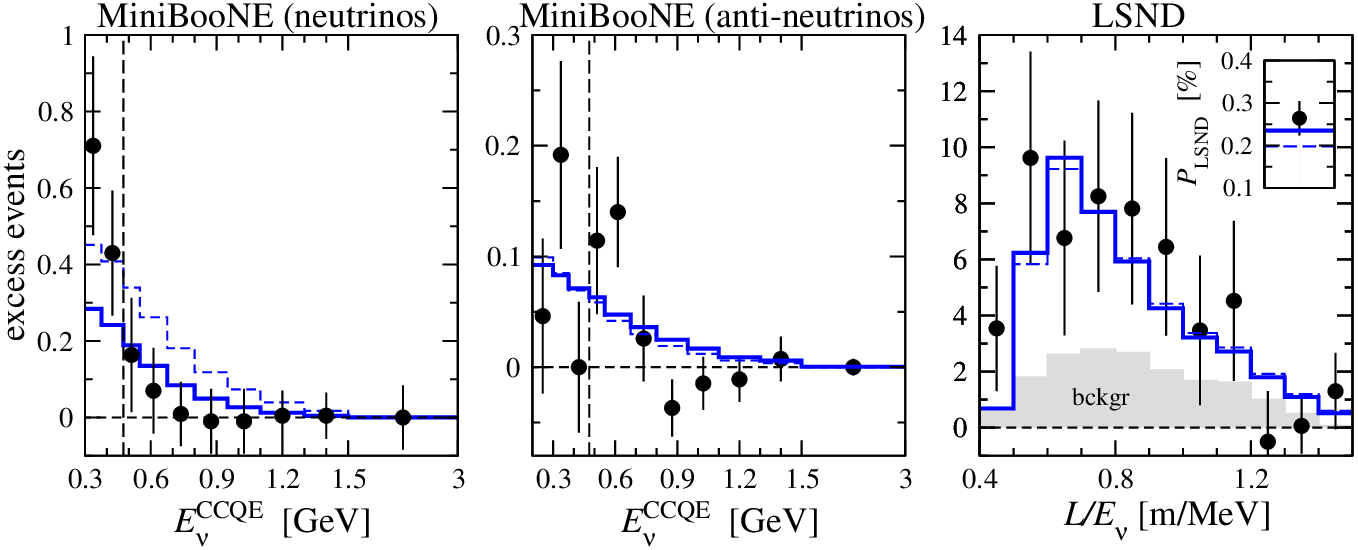}
  \mycaption{\label{fig:spect-app} Predicted event spectra at the best fit
  point to appearance data. Shown are the predictions for MiniBooNE neutrino
  (left), MiniBooNE anti-neutrino (middle), and LSND (right) compared to data.
  Dashed histograms refer to (3+1) oscillations, solid histograms to (3+1)
  NSI. The dashed vertical lines indicate the 475~MeV threshold used for
  MiniBooNE data. For LSND we show the spectrum with free normalisation and
  the value for the total transition probability (inset). The parameter and
  $\chi^2$ values are given in tab.~\ref{tab:fit-params}.}
\end{figure}

First we consider appearance data only, i.e., MiniBooNE neutrino and
anti-neutrino data, LSND, KARMEN, and NOMAD. The best fit parameters and
$\chi^2$ values are given in tab.~\ref{tab:fit-params}, and
fig.~\ref{fig:spect-app} shows the best fit spectra for MiniBooNE and LSND,
where the dashed histograms correspond to (3+1) oscillations and solid
histograms include NSI in addition. The figure illustrates the effect of CP
violation: the best fit occurs at $\delta = 1.5\pi$ (maximal CP violation),
and we observe the suppression of events for MiniBooNE neutrinos while
maintaining the signal for anti-neutrinos in MiniBooNE and LSND.
Let us mention that KARMEN and NOMAD provide an important constraint on the
model, in particular they constrain \bet\ to be small. The $\bet^2$ term in
$P_{\mu e}$ corresponds to a pure NSI effect, independent of energy and
distance, see eq.~(\ref{eq:P1}). Therefore the different value
of $L/E_\nu$ of these experiments cannot be used to circumvent the bounds.
This is one reason why a pure NSI explanation is very difficult, while in our
model we can use the interference terms proportional to Re($\alpha_{\mu e}
\beta_{\mu e}^*$) and Im($\alpha_{\mu e} \beta_{\mu e}^*$) in order to
circumvent the constraint on the constant term.
The relatively large value of $\alp\approx 0.2$ obtained in the fit is a
consequence of the rather small $\Dmq_{41}$, similar to the large mixing
angle obtained for an oscillation fit of MiniBooNE anti-neutrino data only,
see~\cite{AguilarArevalo:2010wv}. Such large amplitudes are excluded by
disappearance data due to eq.~(\ref{eq:factor}), as we will see in the
following.

\begin{table}
\begin{tabular}{l@{\quad}c@{\quad}c@{\quad}c@{\quad}c@{\quad}c}
\hline\hline
Data set & $|\alpha_{\mu e}|$ & $|\beta_{\mu e}|$ & $\delta$ & $\Delta m^2_{41}$ & $\chi^2$/dof \\
\hline
Appearance & 0.2075 &  0.0091 &  $1.5\pi$ &  0.1~eV$^2$ & $33.5/(37-4)$\\
Global     & 0.019  &  0.017  &  $1.3\pi$ &  0.89~eV$^2$ & $107/(115-5)$\\
\hline\hline
\end{tabular}
   \mycaption{\label{tab:fit-params} (3+1) NSI$^c$ best fit parameter and
   $\chi^2$ values for appearance data (LSND, MiniBooNE $\nu$, MiniBooNE
   $\bar\nu$, KARMEN, NOMAD), and the global data. Corresponding event
   spectra are shown in figs.~\ref{fig:spect-app} and
   \ref{fig:spect-global}, respectively. For the global data we have
   $\alpha_e = 0.014$ and $\alpha_\mu = 0.026$ with $\alp^2 = \alpha_e
   \alpha_\mu$.}
\end{table}

\begin{figure}
  \includegraphics[width=\textwidth]{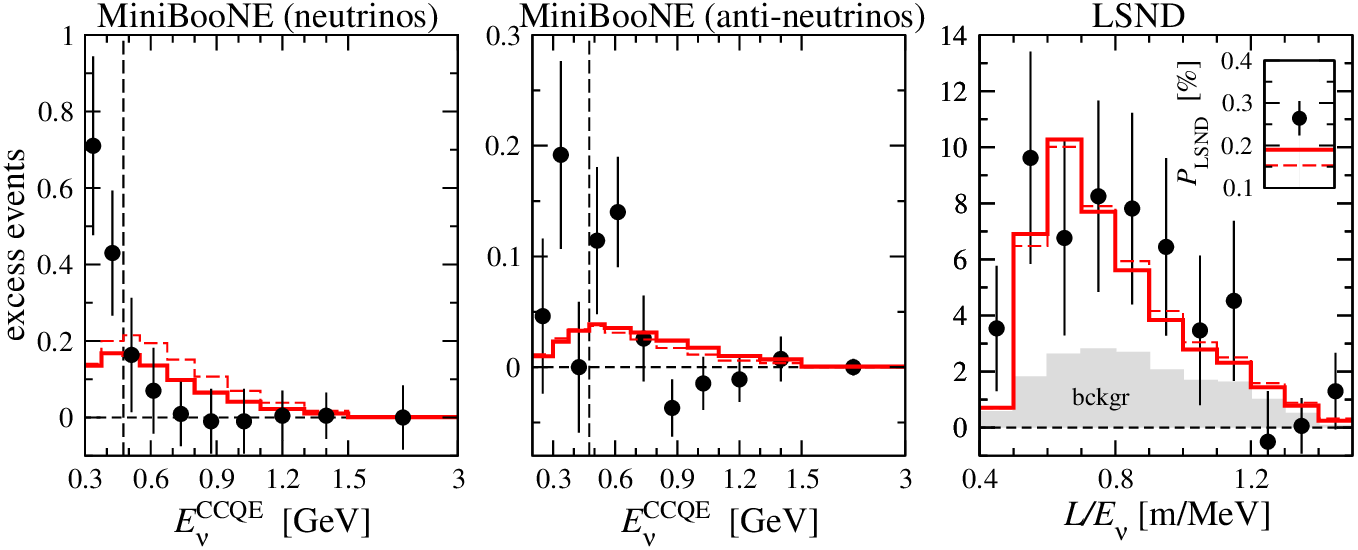}
  \mycaption{\label{fig:spect-global} Predicted event spectra at the best
  fit point to global data. Shown are predictions for MiniBooNE neutrino
  (left), MiniBooNE anti-neutrino (middle), and LSND (right) compared to
  data. Dashed histograms refer to (3+1) oscillations, solid histograms to
  (3+1) NSI$^c$. The dashed vertical lines indicate the 475~MeV threshold used
  for MiniBooNE data. For LSND we show the spectrum with free normalisation
  and the value for the total transition probability (inset). The parameter
  and $\chi^2$ values are given in tab.~\ref{tab:fit-params}.}
\end{figure}

The last row in tab.~\ref{tab:fit-params} gives the results of the global
fit including all data, and the corresponding MiniBooNE and LSND spectra are
shown in fig.~\ref{fig:spect-global}. As mentioned above, disappearance data
constrain \alp\ to be small, and we now obtain \alp\ and \bet\ of similar
order, which maximises the CP effect. The phase $\delta$ is now slightly
non-maximal, but fig.~\ref{fig:spect-global} still shows a relevant
suppression of the events for MiniBooNE neutrino data, which contributes to
the overall improvement of the fit. The constraint from disappearance data
pushes the LSND transition probability to low values, and our best fit
prediction $P_{\rm LSND} = 0.19\%$ is about $1.8\sigma$ away from the
measured value $P_{\rm LSND}^{\rm exp} = (0.264\pm 0.04)\%$ (see inset of
right panel in fig.~\ref{fig:spect-global}).

\begin{figure}
  \includegraphics[width=0.82\textwidth]{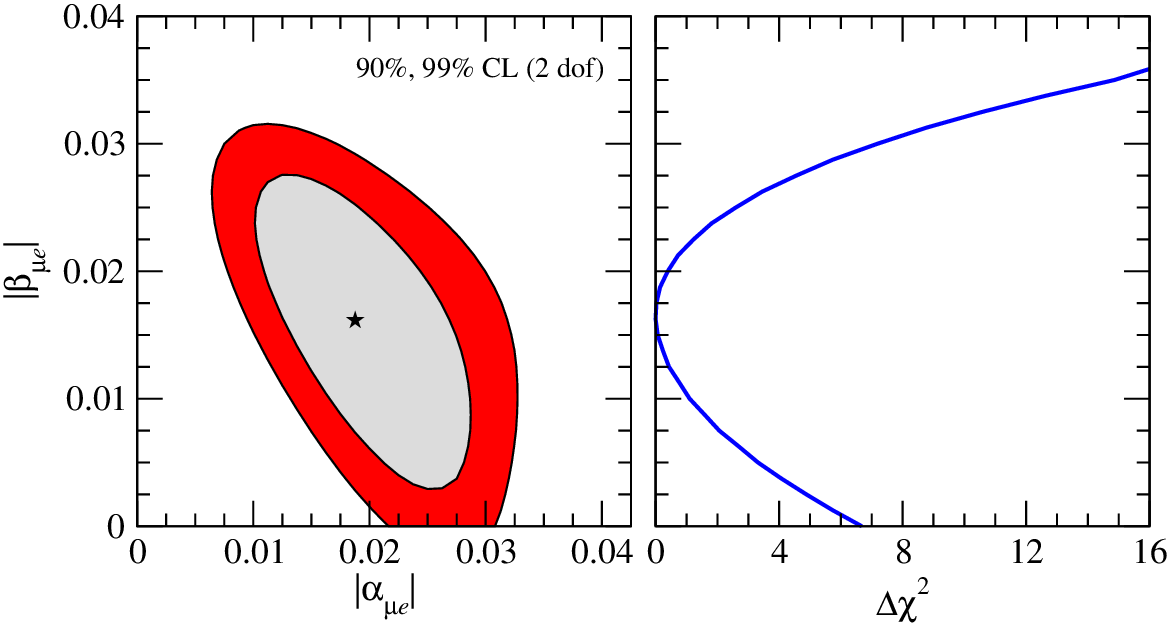}
  \mycaption{\label{fig:alpha-beta} Global fit in the (3+1) NSI$^c$ framework.
  Left: allowed regions projected onto the plane of \alp\ and \bet\ at 90\%
  and 99\%~CL (2~dof). Right: $\Delta\chi^2$ as a function of \bet. We
  minimise over all undisplayed parameters.}
\end{figure}

In fig.~\ref{fig:alpha-beta} we show the allowed regions projected onto the
plane of \alp\ and \bet\ (left) as well as the $\Delta\chi^2$ as a function of
\bet\ (right). The improvement of the fit compared to pure (3+1)
oscillations ($\bet = 0$) is 
\begin{equation}\label{eq:DchiNSI}
\chi^2_\text{min,(3+1)osc} - \chi^2_\text{min,(3+1)NSI$^c$} = 6.9 \qquad\text{(2
dof)}\,,
\end{equation}
where the number of dof corresponds to the two additional parameters of the
(3+1) NSI$^c$ model compared to (3+1) oscillations. Hence, the NSI case is
favoured at 97\%~CL (slightly more than $2\sigma$) compared to the pure
oscillation case. From the right panel of fig.~\ref{fig:alpha-beta} we find
that the allowed interval for \bet\ (1~dof) does not include zero at the
$2.6\sigma$. We conclude that (3+1) NSI provides a significantly better fit
than (3+1) oscillations. However, despite this relative improvement, we
stress that some tension remains in the fit also for (3+1) NSI$^c$, as
discussed below. 

\begin{figure}
  \includegraphics[width=0.55\textwidth]{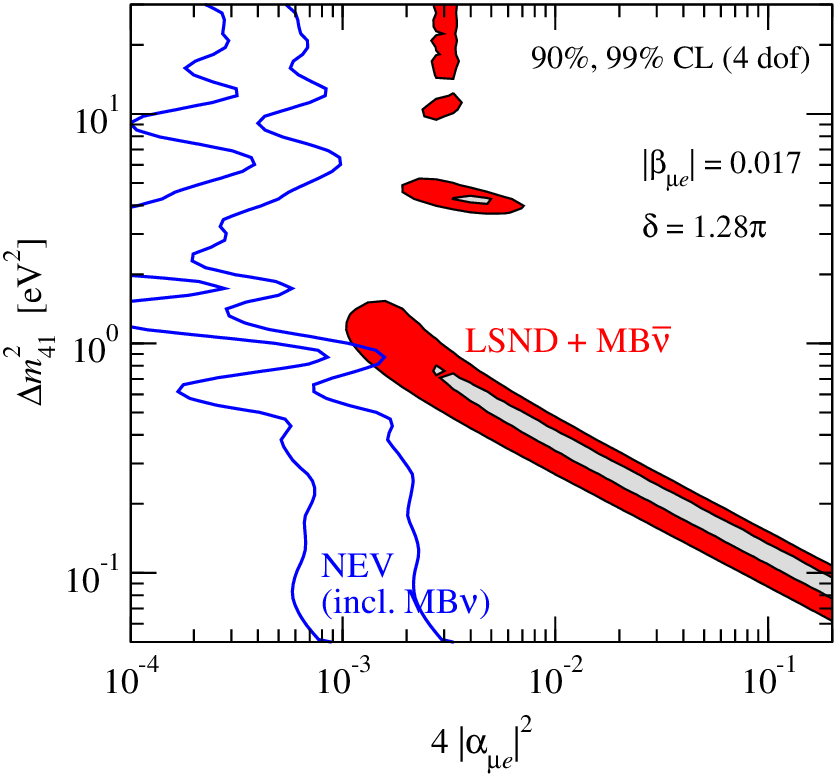}
  \mycaption{\label{fig:ne-vs-ev-nsi} Constraint from no-evidence data (NEV)
  compared to the combined allowed regions from LSND and MiniBooNE $\bar\nu$
  data (shaded) at 90\% and 99\% CL for the (3+1) NSI$^c$ case. The regions
  correspond to sections of the 4-dimensional allowed regions at fixed
  values of $|\beta_{\mu e}|$ and $\delta$ as given in the legend. These
  values correspond to the ones where the NEV constraint and the
  LSND+MB$\bar\nu$ allowed region touch each other (at $\Delta\chi^2 =
  11.7$).}
\end{figure}

Fig.~\ref{fig:ne-vs-ev-nsi} shows the constraint from no-evidence
experiments compared to the allowed region from LSND combined with MiniBooNE
anti-neutrino data. This figure should be compared to
fig.~\ref{fig:ne-vs-ev-osc} for (3+1) oscillations. Note that the parameter
$4\alp^2$ used in fig.~\ref{fig:ne-vs-ev-nsi} corresponds to the effective
mixing angle $\sin^22\theta_{\rm SBL}$ used in fig.~\ref{fig:ne-vs-ev-osc}
in the oscillation limit $\bet\to0$. The regions shown in
fig.~\ref{fig:ne-vs-ev-nsi} are sections of the 4 dimensional volume (in the
space $\alp,\bet,\delta,\Dmq_{41}$) at the fixed values of $|\beta_{\mu e}|$
and $\delta$ for which the two regions start to touch each other. This
happens at a $\Delta \chi^2 = 11.7$, corresponding to 98\%~CL for 4~dof.
Therefore, we observe an overlap of the 99\%~CL regions. Clearly, while some
tension remains in the fit, introducing NSI reduces the disagreement between
evidence and no-evidence data.

\bigskip

Let us now address the question of how to realize the effective parameters
from eq.~(\ref{eq:params_q}) in terms of the fundamental parameters
$U_{\alpha i}$ and $\varepsilon_{\alpha\beta}$ while respecting the
phenomenological bounds. In fact, it
is enough to assume only one single NSI parameter to be different from zero,
namely $\varepsilon_{e\mu}^{ud}$. Then, we neglect the quadratic term in the
normalisation factor $C^X_e$ and the term containing the product of two
small quantities, $\varepsilon_{e\mu}^{ud}$ and $U_{\mu 4}$, in the second
equation in~(\ref{eq:FeFmu}). As a result, we arrive at the following
identification of the parameters relevant to our calculation: 
\be
\alpha_{\mu e}=U_{e4}^{} U_{\mu 4}^*\,,\qquad \beta_{\mu e}=
\varepsilon_{e\mu}^{ud*}\,,\qquad \alpha_{e}=|U_{e4}|^2\,,\qquad 
\alpha_{\mu}=|U_{\mu 4}|^2\,. \label{eq:param} 
\ee 
From tab.~\ref{tab:fit-params} we obtain 
\begin{equation}\label{eq:eps-fit}
|\varepsilon_{e\mu}^{ud}| \approx \bet \approx 0.017 \,
\end{equation}
at the global best fit point. This value is in safe agreement with the
current bound on this parameter from CKM unitarity and lepton universality,
which is of order 0.04~\cite{Biggio:2009nt}. Note that the NOMAD experiment
sets an even stronger bound of about 0.026 on this parameter. Since NOMAD
data is explicitly included in the fit it is clear that our result respects
even this stronger bound. While eq.~(\ref{eq:param}) is just one simple
example, we do not rule out here the possibility that other combinations of 
$U_{\beta 4}$ and $\varepsilon_{\alpha\beta}$ (including $\beta = s$ for the
sterile neutrino) may lead to a similar fit.

One question that is still to be answered is if it is sufficient to have 
only one non-vanishing parameter $\varepsilon_{\alpha\beta}$ in order 
to have physically observable CP violation in neutrino transition 
probabilities. CP violation enters our formulas through the parameter
${\rm Im}\left(U_{e4} U_{\mu 4}^*\varepsilon_{e\mu}^{ud}\right)$. 
It is easy to see that the phase of the expression in the brackets cannot 
be rotated away by a rephasing of the involved fields, so that CP 
violation is physical in the case we consider.%
\footnote{Indeed, $U_{e4}$ can be made real, e.g., by rephasing $\nu_4$, 
and the phase of $U_{\mu 4}$ can be rotated away by rephasing $\mu$.
However, the parameter $\varepsilon_{e\mu}^{ud}$ cannot be made real after 
that because the phases of the $e$ and $\nu_\mu$ fields are already fixed.
Note that a re-phasing of $\mu$ must be accompanied by the corresponding
re-phasing of $\nu_\mu$ due to the standard CC interaction term. One cannot
eliminate the phase of $\varepsilon_{e\mu}^{ud}$ by rephasing the fields of
$u$ or $d$ quarks that enter into eq.~(\ref{eq:eps}) because these fields
enter similarly into both standard and non-standard interactions, and so the
relative phase between these two kinds of terms is unaffected by rephasing
$u$ or $d$.}

\subsection{The general (3+1) NSI model}
\label{sec:NSIg}

Let us now move to the general (3+1) NSI model (NSI$^g$), still assuming
$\varepsilon^S = \varepsilon^D$ for $\nu_\mu$ disappearance experiments, see
eq.~(\ref{eq:source-det}). As discussed in section~\ref{sec:params}, now
LSND and KARMEN data are fitted by the new independent parameters
$|\alpha_{\mu e}^{\rm LK}|$, $|\beta_{\mu e}^{\rm LK}|$, $\delta^{\rm LK}$,
while all other experiments depend on a different set of parameters
$\alpha_e$, $\alpha_\mu$, $|\beta_{\mu e}|$, $\delta$, where the
factorisation $\alp = \sqrt{\alpha_e\alpha_\mu}$ holds. The only parameter
in common between LSND/KARMEN and the rest is $\Dmq_{41}$. Therefore, we
have now 8 parameters in total.

\begin{figure}
  \includegraphics[width=\textwidth]{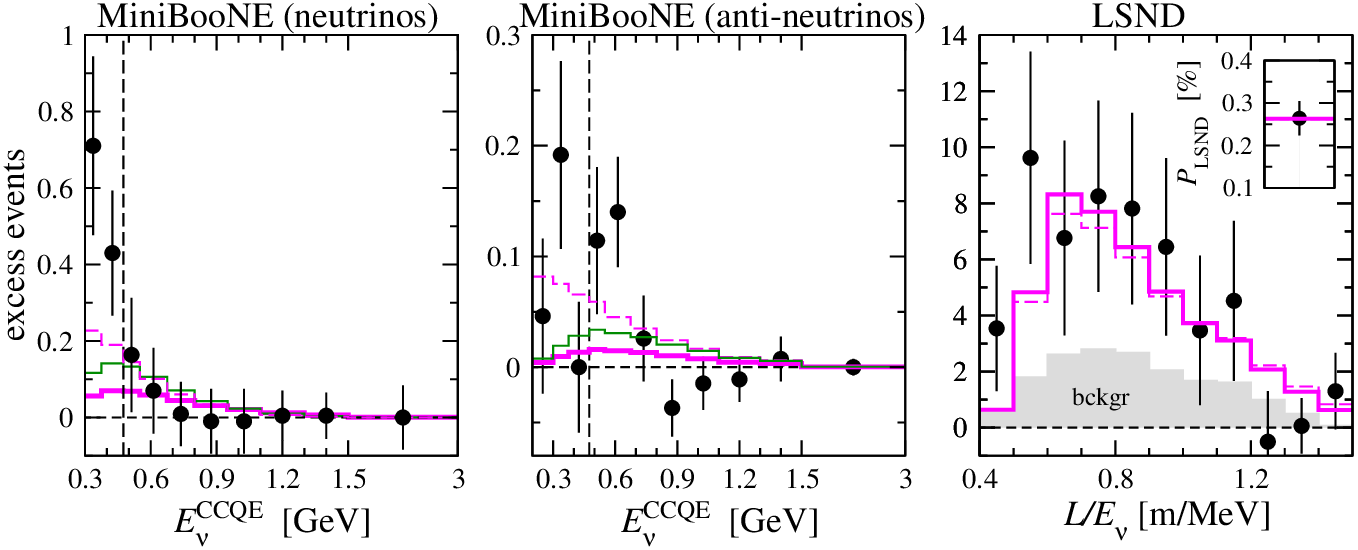}
  \mycaption{\label{fig:spect-unconst} Predicted event spectra at the (3+1)
  NSI$^g$ best fit point to appearance data (dashed) and global data
  (thick-solid).  Shown are predictions for MiniBooNE neutrino (left),
  MiniBooNE anti-neutrino (middle), and LSND (right) compared to data.  The
  thin-solid (green) histograms for MiniBooNE show the global fit result
  without the assumption $\varepsilon^S=\varepsilon^D$ for $\nu_\mu$
  disappearance data (see appendix~\ref{sec:vector}). The dashed vertical
  lines indicate the 475~MeV threshold used for MiniBooNE data. For LSND we
  show the spectrum with free normalisation and the value for the total
  transition probability (inset). The corresponding parameter values are
  given in tab.~\ref{tab:bf-unconst}.}
\end{figure}

In tab.~\ref{tab:bf-unconst} and fig.~\ref{fig:spect-unconst} we show the
result of fits to appearance data only and global data. In the first
case we obtain an excellent description of MiniBooNE and LSND data. We can
invoke CP violation to reconcile MiniBooNE neutrino and anti-neutrino data,
while maintaining an excellent fit to LSND data.\footnote{It turns out that
allowing for NSI also leads to an improvement in the joint LSND + KARMEN
fit. Due to the interference terms between \alpl\ and \betl\ there is more
flexibility in the energy dependence of the transition probability, which
improves the compatibility of the two experiments compared to a pure
oscillation fit by several units in $\chi^2$.} 
However, in the second case, at the global best fit point, we observe
from fig.~\ref{fig:spect-unconst} that the excess in MiniBooNE
anti-neutrino data is not explained. The reason is that the constraint
coming from MiniBooNE neutrino data as well as from the disappearance
experiments is much stronger than the positive signal in MiniBooNE
anti-neutrino data. This is different in the NSI$^c$ case (c.f.,
fig.~\ref{fig:spect-global}), where the signal for MiniBooNE
anti-neutrino data is directly linked to the LSND signal which is
statistically much stronger. Let us mention that should the MiniBooNE
anti-neutrino signal become more significant in the future, a
mechanism similar to the one in the (3+1) NSI$^c$ case (CP violation
due to NSI) can be invoked to explain the excess while satisfying the
bounds.

\begin{table}
\begin{tabular}{l@{\quad}c@{\quad}c@{\quad}c@{\quad}c@{\quad}c@{\quad}c@{\quad}c@{\quad}c}
\hline\hline
Data set & $|\alpha_{\mu e}^{\rm LK}|$ & $|\beta_{\mu e}^{\rm LK}|$ & $\delta^{\rm LK}$
         & $|\alpha_{\mu e}|$ & $|\beta_{\mu e}|$ & $\delta$ & $\Delta m^2_{41}$ & $\chi^2$/dof \\
\hline
Appearance & 0.31 & 0.029 & $0.49\pi$ & 0.15 & 0.011 & $1.5\pi$ & 0.13 eV$^2$
& $29.4/(37 - 7)$ \\
Global     & 0.053 & 0.036 & $0.39\pi$ & 0.010 & 0.013 & $1.2\pi$ & 0.89
            eV$^2$ & $95.4/(115-8)$ \\
\hline\hline
\end{tabular}
   \mycaption{\label{tab:bf-unconst} (3+1) NSI$^g$ best fit parameter and
   $\chi^2$ values for appearance data (LSND, MiniBooNE $\nu$, MiniBooNE
   $\bar\nu$, KARMEN, NOMAD), and the global data. Corresponding event
   spectra are shown in figs.~\ref{fig:spect-unconst}. For the global data we have
   $\alpha_e = 0.010$ and $\alpha_\mu = 0.011$ with $\alp^2 = \alpha_e
   \alpha_\mu$.}
\end{table}

In the global fit we find a minimum $\chi^2$ value of 95.4, which
corresponds to 
\begin{equation}\label{eq:DchiNSIg}
\chi^2_\text{min,(3+1)osc} - \chi^2_\text{min,(3+1)NSI$^g$} = 18.5
\qquad\text{(5 dof)}\,,
\end{equation}
where the number of dof corresponds to the additional 5 new parameters when
extending the (3+1) oscillation scheme to NSI$^g$. The $\Delta\chi^2$ value 
corresponds to 99.76\%~CL. Hence, (3+1) oscillation can be excluded at the
$3\sigma$ level compared to the NSI$^g$ case. In contrast to the NSI$^c$
model, here the tension between appearance and disappearance experiments is
significantly relaxed, since LSND (which provides the main appearance
signal) is decoupled from the disappearance experiments. We will quantify
this in the next section. Let us also mention that there is a nearly
degenerate minimum to the one given in tab.~\ref{tab:bf-unconst} at
$\Dmq_{41} \approx 1.8$~eV$^2$, see fig.~\ref{fig:chisq-Dmq} below.

\begin{figure}
  \includegraphics[width=0.8\textwidth]{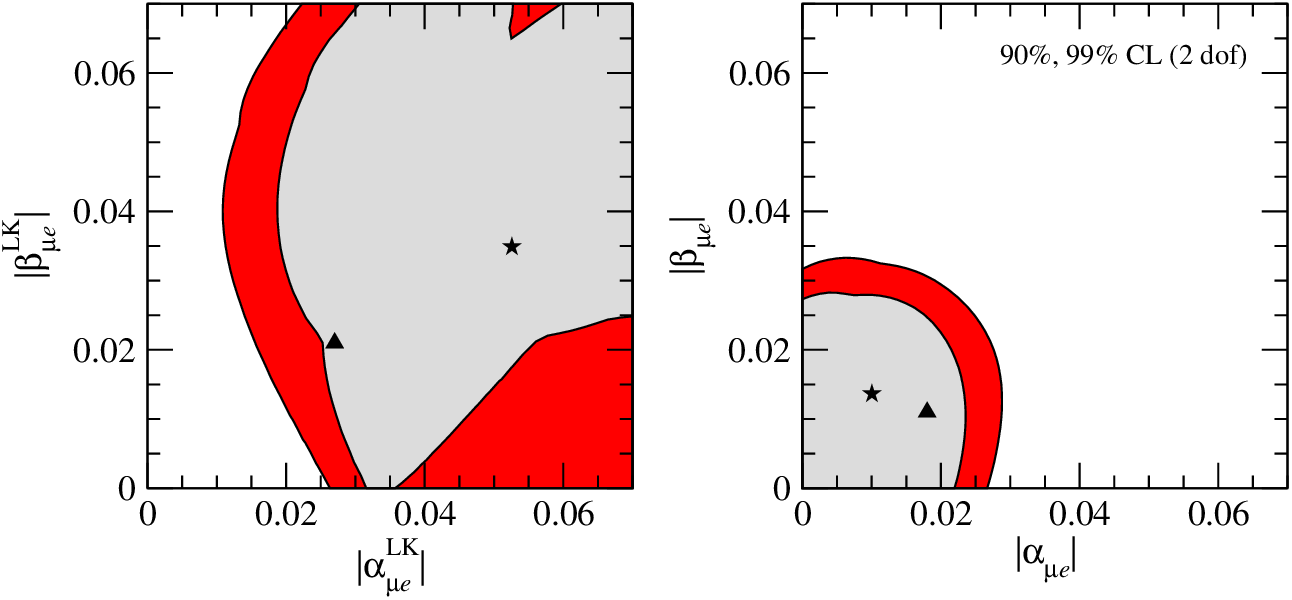}
  \mycaption{\label{fig:alpha-beta-unconst} Global fit in the (3+1) NSI$^g$
  framework. Left: allowed regions projected onto the plane of \alpl\ and
  \betl\ relevant for LSND and KARMEN, right: allowed regions projected onto
  the plane of \alp\ and \bet\ relevant for all the other experiments.
  Regions are shown at 90\% and 99\%~CL (2~dof). We minimise over all
  undisplayed parameters. The stars indicate the global best fit point,
  whereas the triangles correspond to the example from eq.~(\ref{eq:example}).}
\end{figure}

We see from tab.~\ref{tab:bf-unconst} that the global NSI$^g$ best fit point
requires rather large values for the parameters $|\alpl|$ and $|\betl|$ relevant
for LSND and KARMEN. However, the allowed regions shown in
fig.~\ref{fig:alpha-beta-unconst} extend to rather small values even at the
90\%~CL. In the following we provide one possible realization in terms of
fundamental mixing and NSI parameters. It turns out that the main difficulty in
finding $\varepsilon$ parameters within the present bounds is to obtain
$|\alpl| \gtrsim 0.02$ while maintaining $\alp \lesssim 0.02$, compare with
fig.~\ref{fig:alpha-beta-unconst}. 

Let us take the following $\varepsilon$ to be non-zero:
\be
|\varepsilon^{ud}_{\mu s}| \approx 0.05 \,,\qquad
|\varepsilon^{ud}_{e \mu}| \approx 0.011 \,, \qquad
|\varepsilon^{e\nu}_{\mu s}| \approx 0.03 \,,\qquad
|\varepsilon^{e\nu}_{\mu e}| \approx 0.01 \,. \label{eq:eps-fit2}
\ee
Here the superscript $e\nu$ indicates the purely leptonic NSI relevant for
the muon decay. These values are in agreement with the bounds derived
in~\cite{Biggio:2009nt}. Using the fit results and eqs.~(\ref{eq:mue1}),
(\ref{eq:FeFmu}), (\ref{eq:betamue}) we obtain
\bea
\alp &\approx& (|U_{\mu 4}| - |\varepsilon^{ud}_{\mu s}|)|U_{e 4}| 
      \approx 0.018 \,, \nonumber\\
|U_{e 4}| &\approx & 0.116 \,,\qquad |U_{\mu 4}| \approx 0.205 \,,\nonumber\\
\bet &\approx& |\varepsilon^{ud}_{e \mu}| \approx 0.011 \,,\nonumber\\
|\alpl| &\approx& (|U_{\mu 4}| + |\varepsilon^{e\nu}_{\mu s}|)|U_{e 4}| 
      \approx 0.027 \,,\nonumber\\
|\betl| & \approx & |\varepsilon^{ud}_{e \mu}| + |\varepsilon^{\nu e}_{\mu e}|
        \approx 0.021 \,. \label{eq:example}
\eea
Here we have assumed $|U_{s4}| \approx 1$ and neglected terms that are
quadratic in small quantities. Note that we make use of the freedom to
choose the phases of the $\varepsilon$ to suppress \alp\ and enhance
$|\alpl|$. The $\chi^2$ at this point (with $\Dmq_{41} = 0.98$~eV$^2$) is
$\chi^2 = 101.0$, about 5.6 units larger than the best fit point from
tab.~\ref{tab:bf-unconst}, obtained in the phenomenological analysis without
taking into account constraints on NSI parameters. Considering that the
model has 8 parameters, a $\Delta\chi^2 = 5.6$ corresponds to 69\%~CL.
Therefore, the point from eq.~(\ref{eq:example}) is located close to the
$1\sigma$ volume in the 8-dimensional parameter space.\footnote{The value of
$|\varepsilon^{ud}_{\mu s}|$ given in eq.~(\ref{eq:eps-fit2}) is somewhat
large (though in agreement with the bounds from~\cite{Biggio:2009nt}). It is
used to partially cancel the value of $|U_{\mu 4}|$ in \alp. A solution with
$|\varepsilon^{ud}_{\mu s}| = 0$ leads to a fit of similar quality with
$\Delta\chi^2 \approx 7.7$.}

Let us stress that this point is not optimised and should just serve as an
example. Ideally a full analysis should be performed in terms of the
fundamental parameters $U_{\alpha 4}$ and $\varepsilon_{\alpha\beta}$,
taking into account the constraints on the latter. We leave such a fit for
future work and proceed here with the phenomenological analysis in terms of
the $\alpha$ and $\beta$ parameters.

\bigskip

Let us comment now on the relevance of the assumption
eq.~(\ref{eq:source-det}), of equal NSI at production and detection for
$\nu_\mu$ disappearance experiments, which we so far have always adopted.
This assumption can be relaxed by using the fact that vector-like NSI do not
contribute to pion decay. As discussed in appendix~\ref{sec:vector}, this
allows one to decouple the $\nu_\mu$ disappearance data. The results of such a
fit are shown as the thin-solid (green) histogram in
fig.~\ref{fig:spect-unconst}. We find that the predicted spectra for
MiniBooNE are qualitatively very similar to our default NSI$^g$ fit. We
obtain at the best fit point $\chi^2 = 92.7$, to be compared with 95.4 for
the standard NSI$^g$ including the assumption eq.~(\ref{eq:source-det}).
Hence, the improvement of the fit by relaxing this assumption is not
significant. The reason is that even with the assumption $\varepsilon^S =
\varepsilon^D$ for $\nu_\mu$ disappearance experiments, the constraints from
CDHS and atmospheric data are satisfied in the global fit. Therefore,
relaxing this assumption leads only to an insignificant improvement of the
fit. Furthermore, as discussed in appendix~\ref{sec:vector}, this solution
requires relatively large NSI parameters (of order 0.1) and cancellations
between $\varepsilon$'s and elements of the mixing matrix. For these reasons
we do not consider this possibility further and stick to the NSI$^g$
scenario with the assumption eq.~(\ref{eq:source-det}).

\section{Comparison of (3+1) NSI with (3+2) oscillations}
\label{sec:3+2}

\begin{table} 
    \begin{tabular}{l@{\quad}ccc@{\quad}ccc@{\quad}c@{\quad}c@{\quad}c}
	\hline\hline
	& \multicolumn{2}{c}{$|U_{e4} U_{\mu 4}|$} & $\Dmq_{41}$ 
	& \multicolumn{2}{c}{$|U_{e5} U_{\mu 5}|$} & $\Dmq_{51}$  
	& $\delta$ & $\chi^2 / \text{dof}$ 
	\\
	Appearance &
	\multicolumn{2}{c}{0.397} & 0.94 eV$^2$ &
	\multicolumn{2}{c}{0.375} & 1.0 eV$^2$ & 
	1.01$\pi$ & $26.3/(37 - 5)$ 
	\\
	\hline
	& $|U_{e4}|$ & $|U_{\mu 4}|$ & 
	& $|U_{e5}|$ & $|U_{\mu 5}|$ & 
	& &
	\\
	Global data &
	0.10 & 0.15 & 0.47 eV$^2$ &
	0.13 & 0.17 & 0.89 eV$^2$ &
	1.69$\pi$ & $109/(115 - 7)$ 
	\\
	\hline\hline
    \end{tabular}
    \mycaption{\label{tab:3p2}%
      Parameter and $\chi^2$ values of the best fit points in the (3+2)
      oscillation scheme for appearance data from LSND, MiniBooNE $\nu$ and
      $\bar\nu$, KARMEN, NOMAD (upper part), and global data (lower part).}
\end{table}

In this section we present an update of the (3+2) oscillation analysis from
\cite{Maltoni:2007zf} with respect to the anti-neutrino data from MiniBooNE.
In this model appearance and global data depend on 5 and 7 parameters,
respectively, as given in tab.~\ref{tab:3p2}. One physical complex phase
allows for CP violation in $\nu_\mu\to\nu_e$
oscillations~\cite{Karagiorgi:2006jf} and is given by
\begin{equation} 
    \delta \equiv
    \arg\left(U_{e4}^* U_{\mu 4} U_{e5} U_{\mu 5}^* \right) \,. 
\end{equation}

\begin{figure}
  \includegraphics[width=\textwidth]{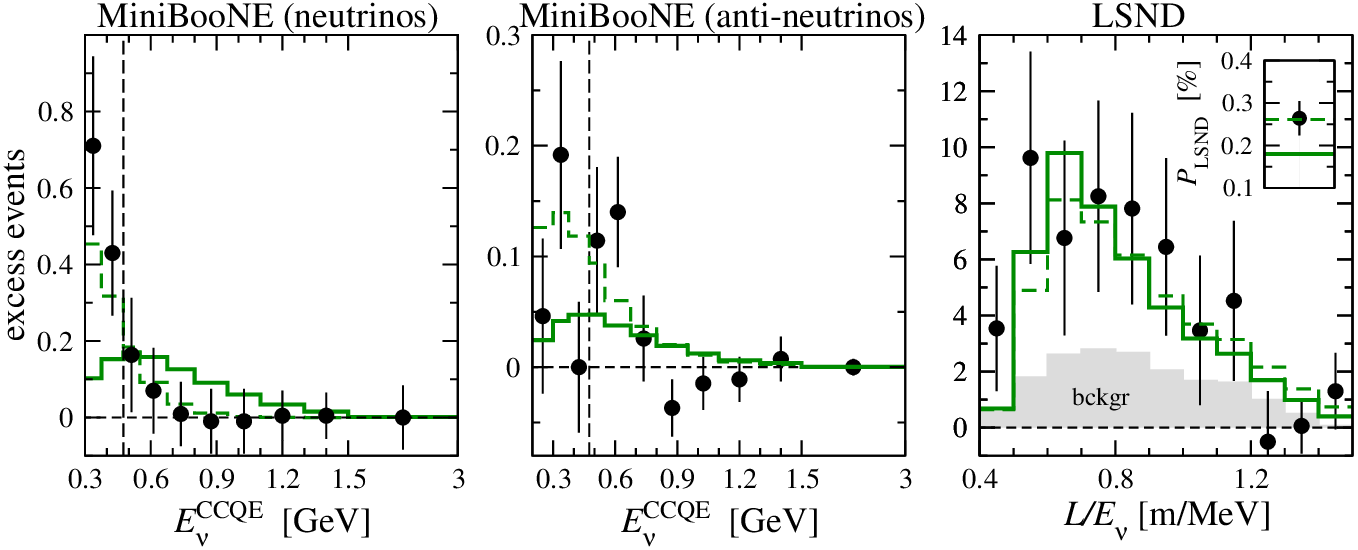}
  \mycaption{\label{fig:spect3p2} Predicted event spectra at the (3+2)
  oscillation best fit point to appearance data (dashed) and global data
  (solid). Shown are predictions for MiniBooNE neutrino (left), MiniBooNE
  anti-neutrino (middle), and LSND (right) compared to data. The dashed
  vertical lines indicate the 475~MeV threshold used for MiniBooNE data. For
  LSND we show the spectrum with free normalisation and the value for the
  total transition probability (inset). The corresponding parameter values
  are given in tab.~\ref{tab:3p2}.}
\end{figure}

The best fit points for appearance and global data are given in 
tab.~\ref{tab:3p2} and the corresponding event spectra for MiniBooNE and
LSND are shown in fig.~\ref{fig:spect3p2}. Considering appearance data only,
a very good fit is obtained for neutrino and anti-neutrino data thanks to CP
violation and even the MiniBooNE $\nu$ and $\bar\nu$ low
energy data can be explained due to some cancellations between terms
involving the two mass-squared differences and $\delta \approx \pi$, as
discussed in \cite{Maltoni:2007zf}. However, rather large values of the
amplitudes $|U_{e4} U_{\mu 4}|$ and $|U_{e5} U_{\mu 5}|$ are needed, in
disagreement with constraints from disappearance data.

In the global analysis some tension is visible in the spectra shown in
fig.~\ref{fig:spect3p2}. The predicted event rate for MiniBooNE neutrino
data is somewhat too high and the probability for LSND is somewhat low:
$P_\mathrm{LSND} = 0.18\%$, similar to the (3+1) NSI$^c$ case. Furthermore, it is
not possible to explain the low energy data in MiniBooNE below 475~MeV.
The relative improvement of the fit due to introducing the second sterile
neutrino is 
\begin{equation}\label{eq:Dchi3p2}
\chi^2_\text{min,(3+1)osc} - \chi^2_\text{min,(3+2)osc} = 5.0 \qquad\text{(4 dof)}\,,
\end{equation}
where the number of dof corresponds to the additional 4 parameters
introduced by moving from (3+1) to (3+2). The value given in
eq.~(\ref{eq:Dchi3p2}) corresponds to 71\%~CL, to be compared to 97\%~CL for
NSI$^c$ (eq.~(\ref{eq:DchiNSI})) and 99.76\%~CL for NSI$^g$
(eq.~(\ref{eq:DchiNSIg})). We conclude that both versions of the NSI model
(but especially the unconstrained case) offer significantly more improvement
of the global fit compared to the (3+2) oscillation scheme, when considering
the $\chi^2$ gain per new parameter. 
The previous (3+2) value obtained in~\cite{Maltoni:2007zf} without
MiniBooNE anti-neutrino data was 81\%~CL ($\Delta \chi^2=6.1$). Therefore,
we find that the relative improvement of the fit when moving from (3+1) to
(3+2) oscillations is smaller with the new global data than without MiniBooNE
$\bar\nu$ data.

\begin{figure}
  \includegraphics[width=0.6\textwidth]{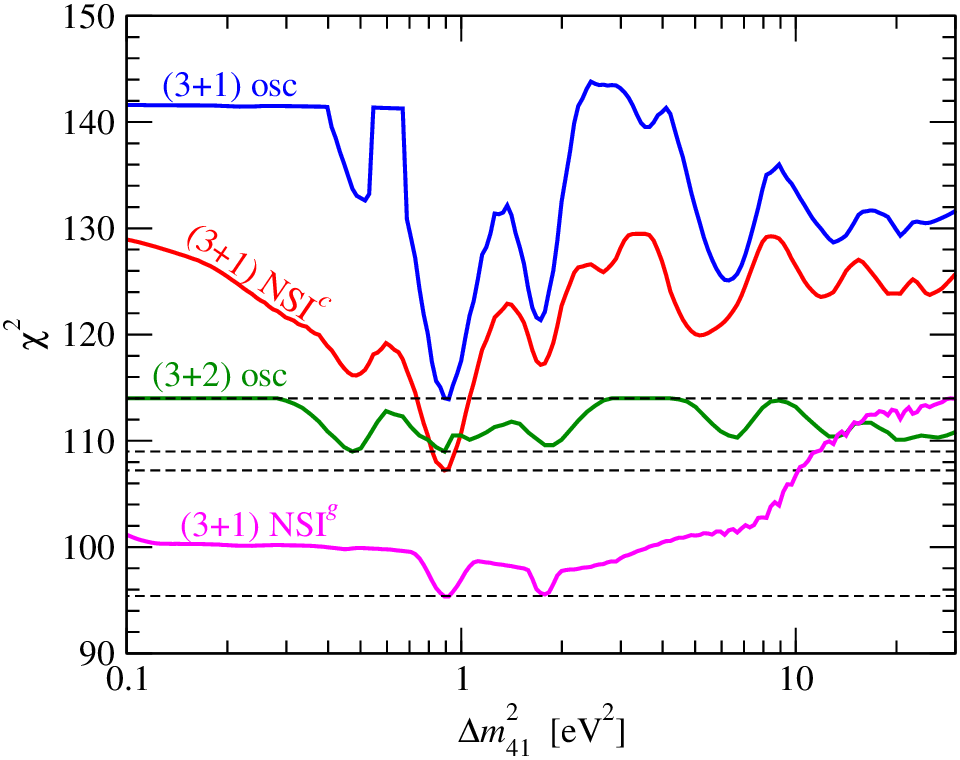}
  \mycaption{\label{fig:chisq-Dmq} $\chi^2$ of global data as a function of
  $\Dmq_{41}$ for the (3+1) oscillation, (3+1) NSI$^c$, (3+1) NSI$^g$, and
  (3+2) oscillation models. In each case we minimise with respect to all
  parameters except $\Dmq_{41}$. The horizontal dashed lines indicate the
  global minima.}
\end{figure}

The $\chi^2$ profiles as a function of $\Dmq_{41}$ for the four
models---(3+1) oscillation, (3+1) NSI$^c$, (3+1) NSI$^g$, and (3+2)
oscillation---are shown in fig.~\ref{fig:chisq-Dmq}. In this plot we minimise
with respect to all parameters except $\Dmq_{41}$. Note that the maximal
$\chi^2$ for (3+2) is given by the $\chi^2$ minimum in (3+1), since the
(3+1) solution with $\Dmq_{51}$ is always available for any value of
$\Dmq_{41}$. Note also that there are two degenerate minima for (3+2) at
$\Dmq_{41} = 0.47$~eV$^2$ and 0.89~eV$^2$ corresponding just to a relabeling
of the mass-squared differences, compare with tab.~\ref{tab:3p2}. The relative
improvement of the three models NSI$^c$, NSI$^g$, and (3+2) with respect to
the (3+1) oscillation case is summarised in the lower part of
tab.~\ref{tab:PG}.

\begin{table}[t] \centering
    \begin{tabular}{l@{\qquad}c@{\quad}c@{\qquad}c@{\quad}c@{\qquad}c@{\quad}c}
	\hline\hline
	& \multicolumn{2}{c}{(3+1) NSI$^c$} 
	& \multicolumn{2}{c}{(3+1) NSI$^g$} 
	& \multicolumn{2}{c}{(3+2) oscillations}
	\\
	\hline
	& $\chi^2_{\rm PG} / \text{dof}$ & PG prob.
	& $\chi^2_{\rm PG} / \text{dof}$ & PG prob.
	& $\chi^2_{\rm PG} / \text{dof}$ & PG prob.
	\\
	Evid. vs no-evid. &
	23.3/4 & $1.1\times 10^{-4}$ & & & 26.9/5 & $6\times 10^{-5}$
	\\
	Appear. vs disapp. &
	11.5/2 & $3\times 10^{-3}$ &
	 3.8/2 & 15\% &
	21.7/4 & $2.3\times 10^{-4}$
	\\
	\hline
	& $\Delta\chi^2/ \text{dof}$ & CL 
	& $\Delta\chi^2/ \text{dof}$ & CL 
	& $\Delta\chi^2/ \text{dof}$ & CL 
	\\
	Fit wrt (3+1) osc. &
	6.9/2 & 97\% & 18.5/5 & 99.76\% & 5.0/4 & 71\%
	\\
	\hline\hline
    \end{tabular}
    \mycaption{\label{tab:PG} Comparison of (3+1) NSI and (3+2)
    oscillations. We show the compatibility of LSND + MiniBooNE
    anti-neutrino data (``Evid.'') compared to the rest of the global data
    (``no-evid.''), and the appearance versus disappearance experiments.
    $\chi^2_{\rm PG}$, number of dof, and the corresponding probability are
    given. The lower part of the table shows the improvement with respect to
    the (3+1) pure oscillation case. We give the improvement in $\chi^2$,
    where the dof corresponds to the number of additional parameters.}
\end{table}

Further insight in the quality of the fit can be obtained by evaluating the
compatibility of different data sets with the so-called parameter
goodness-of-fit (PG)~\cite{Maltoni:2003cu}. It is based on the $\chi^2$
function $\chi^2_\text{PG} = \chi^2_\text{tot,min} - \sum_i
\chi^2_{i,\text{min}}$, where $\chi^2_\text{tot,min}$ is the $\chi^2$
minimum of all data sets combined and $\chi^2_{i,\text{min}}$ is the minimum
of the data set $i$. This $\chi^2$ function measures the ``price'' one has
to pay by the combination of the data sets compared to fitting them
independently. It should be evaluated for the number of dof corresponding to
the number of parameters in common to the data sets,
see~\cite{Maltoni:2003cu} for a precise definition. 

In tab.~\ref{tab:PG} we show the results of such an analysis, testing the
compatibility of evidence versus no-evidence data (similar as shown in
fig.~\ref{fig:ne-vs-ev-nsi}) and appearance versus disappearance data. These
results indicate that in (3+2) as well as in (3+1) NSI$^c$ significant tension
remains between various data sets. 
For the general NSI model NSI$^g$, however, we find excellent agreement between
appearance and disappearance data. In this case LSND is decoupled from the
disappearance experiments and this tension is completely resolved. The
evidence in MiniBooNE anti-neutrino data is not strong enough yet to show
up as significant tension in the PG test. We do not perform the evidence
versus no-evidence test for NSI$^g$, since it makes no sense to add LSND and
MiniBooNE anti-neutrino data here because, as we have seen in
fig.~\ref{fig:spect-unconst}, the MiniBooNE $\bar\nu$ excess is not explained
in the global fit in this model.

\section{Summary and discussion}
\label{sec:summary}

Recent MiniBooNE anti-neutrino data indicate an excess of $\bar\nu_e$
events, in agreement with the LSND evidence for $\bar\nu_\mu\to\bar\nu_e$
transitions. It is known that oscillations with a single sterile neutrino at
the eV scale are not sufficient to explain the global data.  We have
investigated the possibility that in addition to a sterile neutrino there
are some non-standard neutrino interactions (NSI), beyond the Standard Model
weak interactions. Since matter effects are tiny for the short baselines
relevant here, we considered charged-current type NSI in the neutrino source
and detector, parametrised by $\varepsilon_{\alpha\beta}$ as defined in
eq.~(\ref{eq:eps}). Thanks to the interference between NSI effects and
oscillations with $\Dmq_{41} \sim 1$~eV$^2$ we obtain CP violation, even in
the presence of only one mass scale. This effect is used to reconcile the
indication for $\bar\nu_\mu\to\bar\nu_e$ in anti-neutrino experiments (LSND
and MiniBooNE) with the absence of a signal in MiniBooNE neutrino data. 

We have presented a general parameterisation of the relevant transition and
survival probabilities in the presence of oscillations (within the one-mass
scale approximation) and NSI, and we have identified particular combinations
of mixing matrix elements $U_{\alpha 4}$ and NSI parameters
$\varepsilon_{\alpha\beta}$ entering in the probabilities. This drastically
reduces the number of independent parameters and allows us to perform a
general fit to global short-baseline data.

We have considered two versions of the (3+1) NSI model. In the general case
(denoted NSI$^g$) we make use of the fact that the neutrino production
mechanism in LSND (and in KARMEN) is muon decay (purely leptonic), whereas
in all other experiments neutrino production and detection are
semi-leptonic, involving transitions between $u$ and $d$ quarks. Therefore,
in the presence of suitable NSI parameters we can decouple the transition
probabilities in LSND and KARMEN from the rest of the data. In this case we
obtain an excellent fit to the global data and the tension between
appearance and disappearance experiments is resolved. Let us mention that in
this case MiniBooNE does not provide a direct test of LSND, since different
combinations of parameters are relevant for them. Also, in the global fit
the excess observed in MiniBooNE anti-neutrino data is not reproduced. 

For the second version of the (3+1) NSI model we adopt the assumption that
NSI involving the charged muon can be neglected. In this case exactly the
same NSI parameters are relevant for LSND and KARMEN as for all other
experiments. In this constrained model (NSI$^c$) we make use of the CP
violation due to NSI--oscillation interference to reconcile neutrino and
anti-neutrino data. We have shown that in the NSI$^c$ model there is a
factorisation between appearance and disappearance amplitudes, similar to
that in the (3+1) oscillation scheme. Therefore, it is more difficult to
satisfy constraints from disappearance experiments and some tension is left
in the fit. However, also this model provides significant improvement of the
global fit compared to the pure oscillation case. 

We have presented the results of our fits in terms of effective parameters,
representing the specific combinations of NSI parameters entering in the
transition probabilities. However, for both cases, NSI$^c$ and NSI$^g$, we
have provided also examples of how to realise the required parameters in
terms of the fundamental mixing and NSI parameters. We have shown that 
values in safe agreement with bounds on the various $\varepsilon$'s can be
found to realise our fits. The examples given in eqs.~(\ref{eq:eps-fit}) and
(\ref{eq:eps-fit2}) require $\varepsilon$'s of order a few$\,\times 10^{-2}$.

We have compared the quality of the (3+1) NSI fits to an updated fit in the
(3+2) oscillation scheme, which also allows for CP violation due to the
presence of two relevant mass scales. Similarly to (3+1) NSI, in (3+2) the
appearance experiments can be described very well. However, we confirm
previous results that for (3+2) oscillations significant tension remains in
the global fit between appearance and disappearance experiments. The
improvement of (3+2) compared to (3+1) is not significant, in terms of
$\chi^2$ gain per new parameter. 
Let us mention also that in none of the scenarios considered here we can
explain the MiniBooNE low energy excess of events when disappearance data
are taken into account. Therefore, we follow the strategy of the MiniBooNE
collaboration and exclude the data below 475~MeV from the analysis, relying
on a separate explanation for this anomaly.

The predictions of our model for future experiments depend on the detailed
realization in terms of mixing and NSI parameters. In general one may expect
some signals in searches for deviations from the standard three-flavour
oscillation picture in both respects, sterile neutrino oscillations as well
as NSI. Several proposals to search for sterile neutrinos at the eV scale
have been presented recently, see for example~\cite{Donini:2008wz,
Giunti:2009en, Baibussinov:2009tx, Agarwalla:2009em, Meloni:2010zr,
Agarwalla:2010zu}. In~\cite{Hamann:2010bk} implications of sterile neutrinos
for latest cosmological data have been investigated. Recent studies on NSI
in the context of upcoming and far future experiments can be found, e.g.,
in~\cite{Kopp:2007ne, Ohlsson:2008gx, Kopp:2008ds, Meloni:2009cg}.

A specific prediction of our scenario are zero-distance effects in
appearance searches~\cite{Langacker:1988up, FernandezMartinez:2007ms,
Meloni:2009cg}, since our solutions all include some non-vanishing value of
the parameter \bet. This parameter induces a non-zero transition probability
even at zero distance from the neutrino source, see eq.~(\ref{eq:P1}).
Hence, the observation of an energy independent appearance probability at
very short distances is a characteristic signature from this kind of models.
The idea presented in~\cite{Agarwalla:2010zu} could be particularly useful
to search for this effect, since it would allow to map out the $E_\nu$ and
$L$ dependence of a $\bar\nu_e$ appearance signal.

Our model may also provide a signature at the LHC. Typically, realising
CC-like interactions as the ones from eq.~(\ref{eq:eps}) require a charged
particle as mediator. The NSI parameters $\varepsilon$ measure the strength
of the new interactions relative to the standard weak interaction strength
set by $G_F$. Therefore, from our fit results, $\varepsilon \sim 0.01$, one
expects that the mass of a mediator for a dimension-6 operator should be
roughly one order of magnitude larger than the $W$ boson mass. Hence, one
might expect charged particles to show up at the TeV scale, with good
prospects to be observed at LHC. Let us mention, however, that the results
of~\cite{Antusch:2008tz,Gavela:2008ra} suggest that NSI at the level
of 0.01 are difficult to obtain from dimension-6 operators without being in
conflict with bounds on charged-lepton processes. As discussed there, a
possibility to obtain such large NSI would be to go to dimension-8
operators and allow for some fine tuning.

\subsection*{Acknowledgements}

We would like to thank K.S.\ Babu and W.C.\ Louis for stimulating
discussions and M.\ Blennow, E.\ Fernandez-Martinez, and J.\ Kopp for useful
correspondence.
We acknowledge the financial support of the European Community under the
European Commission Framework Programme 7 Design Study EUROnu, Project
Number 212372. The EC is not liable for any use that may be made of the
information contained herein.
This work is partly supported by the Transregio Sonderforschungsbereich TR27
``Neutrinos and Beyond'' der Deutschen Forschungsgemeinschaft.

\appendix

\section{$\varepsilon^S \neq \varepsilon^D$ in $\nu_\mu$ disappearance
experiments}
\label{sec:vector}

As discussed in section~\ref{sec:params}, the fact that pions couple only to
the axial-vector current may lead to the situation that different NSI
contribute at neutrino production and detection in $\nu_\mu$ disappearance
experiments. In particular, vector-like NSI, $\varepsilon^{(V)} =
\varepsilon^{(R)} + \varepsilon^{(L)}$, will only contribute in the
detection process but not at production. This effectively introduces new
independent parameters which decouple also the $\nu_\mu$ disappearance
experiments. 

Let us introduce the following abbreviations for the various production and
detection processes:
\be\label{eq:a1}
\begin{array}{c@{\::\:}l}
\mu   & \text{muon decay} \\
\pi   & \text{pion decay} \\
Ne    & \text{neutrino--nucleus CC interaction involving an electron}\\
N\mu  & \text{neutrino--nucleus CC interaction involving a muon}
\end{array}
\ee
Relaxing now the assumption eq.~(\ref{eq:source-det}) we have the
following set of parameters relevant for the various experiments (in
addition to the common $\Dmq_{41}$):
\be\label{eq:a2}
\begin{array}{l@{\quad}l@{\quad}l}
\text{LSND/KARMEN:} & \alpha_{\mu e} = F^{\mu}_{\mu 4} F^{Ne*}_{e4} \,,
                    & \beta_{\mu e} = \sum_i F^{\mu}_{\mu i} F^{Ne*}_{ei} \\
\text{MiniBooNE/NOMAD:} 
                    & \alpha_{\mu e} = F^{\pi}_{\mu 4} F^{Ne*}_{e4} \,,
                    & \beta_{\mu e} = \sum_i F^{\pi}_{\mu i} F^{Ne*}_{ei} \\
\text{reactor:}     & \alpha_{e e} = |F^{Ne}_{e4}|^2 \,,
                    & \beta_{e e} = 1 \\
\text{CDHS/atmospheric:}  
                    & \alpha_{\mu\mu} = F^{\pi}_{\mu 4} F^{N\mu *}_{\mu 4} \,,
                    & \beta_{\mu\mu} = \sum_i F^{\pi}_{\mu i} F^{N\mu *}_{\mu i} 
\end{array}
\ee
where the $F_{\alpha i}$ are defined in eq.~(\ref{eq:F1}). Now it is possible
to have $P_{\mu\mu} = 1$ while allowing for a non-zero transition
probability in MiniBooNE, which however, requires some cancellation between
NSI parameters and elements of the mixing matrix. For example, one can take
\be 
F^{N\mu}_{\mu 4} \approx 
  U_{\mu 4}^* + \varepsilon^{N\mu}_{\mu s} U_{\mu s}^* \approx 0 
  \,,\qquad 
\varepsilon^{\pi}_{\mu s} \approx 0 \,. \ee
This implies $\alpha_{\mu\mu} \approx 0$ and $\beta_{\mu\mu} \approx 1$ and
therefore $P_{\mu\mu} \approx 1$, as required by the data from CDHS and
atmospheric neutrinos. On the other hand, $F^\pi_{\mu4} \approx U_{\mu 4}^*$
and we can have $P_{\mu e} > 0$ for MiniBooNE, including the possibility of
CP violation. 

The results of such a fit for MiniBooNE are shown as thin-solid (green)
histograms in fig.~\ref{fig:spect-unconst}, which are qualitatively very
similar to our default NSI$^g$ fit. Reactor experiments still constrain the
value of $|F^{Ne}_{e4}|$ to be small, which excludes small values of
$\Dmq_{41}$ where a better fit to the MiniBooNE spectrum would be possible
(such as for example for the appearance data only fit shown in
fig.~\ref{fig:spect-unconst}), and we find $\Dmq_{41} \simeq 0.9$~eV$^2$ at
the best fit point. The spectral shape of the signal for such values of
$\Dmq_{41}$ does not allow for a better fit of MiniBooNE data even without
the constraint from $\nu_\mu$ disappearance. 
Decoupling the $\nu_\mu$ disappearance data by setting $P_{\mu\mu}=1$, we
find a best fit point with $\chi^2 = 92.7$, to be compared with 95.4 for the
standard NSI$^g$ including the assumption eq.~(\ref{eq:source-det}). Hence,
the improvement of the fit by relaxing this assumption is not significant.
The reason is that, despite this assumption, there is already very good
agreement between appearance and disappearance data, as discussed in
section~\ref{sec:3+2}. Even with the assumption $\varepsilon^S =
\varepsilon^D$ for $\nu_\mu$ disappearance experiments, the constraints from
CDHS and atmospheric data are satisfied in the global fit. Therefore,
relaxing this assumption leads only to an insignificant improvement of the
fit.

Furthermore, as mentioned above, decoupling $\nu_\mu$ disappearance data
requires some unpleasant cancellation. At the best fit point shown in
fig.~\ref{fig:spect-unconst} we find $U_{\mu 4} \approx 0.26$. Therefore, to
cancel this in $F^{N\mu}_{\mu 4}$ one needs $\varepsilon^{N\mu}_{\mu s}$ of
the same order. Note that the constraints on $\varepsilon_{\mu \alpha}$ at
the level of a few percent come from pion decay
processes~\cite{Biggio:2009nt}, which only apply to axial-vector-like NSI,
while here we need vector-like NSI precisely to avoid the contribution to
pion decay. Therefore, such large NSI might be phenomenologically viable,
though still uncomfortably large. Moreover, there is no reason why they
should have values such that they cancel against $U_{\mu 4}$, since these
are two unrelated quantities. Hence, together with the observation that the
improvement of the fit is not significant, this motivates us to stick in the
main text to the assumption eq.~(\ref{eq:source-det}) in order to simplify
the analysis.

\section{MINOS results and NSI}
\label{sec:minos}

Let us comment briefly on the possibility to apply NSI in source and
detector as an explanation for recent MINOS results, which indicate a slight
difference between neutrino and anti-neutrino data \cite{minos:an}, where we
restrict ourselves here to an effective two-flavour framework in the
$\mu-\tau$ sector.
The mentioned results from MINOS are based on data from $\nu_\mu$ and
$\bar\nu_\mu$ disappearance searches.  Our formalism allows for CP violation
in the survival probability provided NSI are different at neutrino
production and detection. Hence, one can use the same mechanism as discussed
in appendix~\ref{sec:vector}. Since pions couple only to the axial-vector
current, vector-like NSI will contribute only at neutrino detection but not
at the neutrino source, which is pion decay in MINOS~\cite{joachim}. Note
that the one mass scale dominance approximation used in this work applies
approximately also for MINOS, though for $\Delta m^2_{31}$. We find that
relatively large values of $\varepsilon$ are needed, of order 0.1, while at
the same time the combined fit of MINOS neutrino and anti-neutrino data does
improve only by about 2.5 units in $\chi^2$.  

We have also considered the possibility to use NC-like NSI in the context of
MINOS~\cite{Engelhardt:2010dx, Mann:2010jz, Heeck:2010pg}, which would
induce a non-standard matter effect in the $\mu-\tau$ sector, and which
therefore could lead to a difference between neutrino and anti-neutrino
results. We find that $(i)$ diagonal NSI $\varepsilon^{\rm NC}_{\mu\mu}, 
\varepsilon^{\rm NC}_{\tau\tau} \lesssim 0.5$ have a negligibly small effect
on the fit, and $(ii)$ off-diagonal NSI $\varepsilon^{\rm NC}_{\mu\tau}$ of
the order 0.2 lead to an improvement of the combined fit of about 2.5 units
in $\chi^2$. These values for $\varepsilon^{\rm NC}_{\mu\mu}$,
$\varepsilon^{\rm NC}_{\mu\tau}$, $\varepsilon^{\rm NC}_{\tau\tau}$ are
about one order of magnitude larger than the bounds from atmospheric
neutrino data~\cite{GonzalezGarcia:2004wg}. We conclude that neither CC nor
NC type NSI in the $\mu-\tau$ sector provide a viable explanation for a
possible deviation of neutrino and anti-neutrino results from present MINOS
data. The question whether this conclusions holds also in a more general
three-flavour (or maybe four-flavour) framework is left for future work, see
also~\cite{Friedland:2006pi, Blennow:2007pu}.

\bibliographystyle{my-h-physrev}
\bibliography{./nsi}

\end{document}